# Dual-Comb Hyperspectral Digital Holography

Edoardo Vicentini [1-3], Zhenhai Wang [1,4], Kasper Van Gasse [1,5], Theodor W. Hänsch [1,6], Nathalie Picqué [1,*]

[1] Max-Planck Institute of Quantum Optics, Hans-Kopfermann-Straße 1, 85748, Garching, Germany
[2] Dipartimento di Fisica, Politecnico di Milano, Piazza L. Da Vinci, 32 20133 Milano, Italy
[3] Istituto di Fotonica e Nanotecnologie, Consiglio Nazionale delle Ricerche, Piazza L. Da Vinci, 32 20133 Milano, Italy
[4] Center for Combustion Energy, Department of Energy and Power Engineering, Tsinghua University, 100084 Beijing, China
[5] Photonics Research Group, INTEC, Ghent University - imec, Technologiepark-Zwijnaarde 126, 9052 Ghent, Belgium
[6] Ludwig-Maximilian University of Munich, Faculty of Physics, Schellingstr. 4/III, 80799, München, Germany
*Corresponding author: nathalie.picque@mpq.mpg.de

**Holography [1] has always held special appeal, for it is able to record and display spatial information in three dimensions [2-10]. Here, we show how to augment the capabilities of digital holography [11, 12] by using a large number of narrow laser lines at precisely-defined optical frequencies simultaneously. Using an interferometer based on two frequency combs [13-15] of slightly different repetition frequencies and a lens-less camera sensor, we record time-varying spatial interference patterns that generate spectral hypercubes of complex holograms, revealing, for each comb line frequency, amplitudes and phases of scattered wave-fields. Unlike with previous multi-color holography and low-coherence holography (including with a frequency comb [16] ), the unique synergy of broad spectral bandwidth and high temporal coherence in dual-comb holography opens up novel optical diagnostics, such as precise dimensional metrology over large distances without interferometric phase ambiguity, or hyperspectral 3-dimensional imaging with high spectral resolving power, as we illustrate by molecule-selective imaging of an absorbing gas.**

With a set-up which seemingly involves merely the same simple hardware as that harnessed for dual-comb hyperspectral imaging [17-20] - two frequency-comb generators and a detector matrix-, we take a new route to lens-less scan-free 3D-profiling, suited to dimensional metrology of macroscopic and microscopic objects and to advanced optical sensing that combines spatial and spectral information.

In our technique of dual-comb holography (Fig.1), a frequency-comb generator, called object-comb generator, emits a train of pulses at a repetition frequency $f_{rep}$. Its beam is either transmitted through or reflected by a three-dimensional object. The light scattered by the object is combined on a beam-splitter with the beam of a second frequency-comb generator (called reference), of slightly different repetition frequency $f_{rep}+\delta f_{rep}$. The two trains of pulse wavefronts interfere at a fast lens-less detector matrix and their interference signal at each matrix pixel is sampled as a function of time over a duration longer than $1/\delta f_{rep}$, while full interferometric coherence is maintained. The asynchrony of the two pulse trains enables interference between the object and reference waves over a range of optical delays limited by the mutual coherence time of the interferometer only, which can be longer than one second. However, as the interferometric signal recurs with an optical-delay period of $1/f_{rep}$, the axial ambiguity range is $c/f_{rep}$, where $c$ is the speed of light.

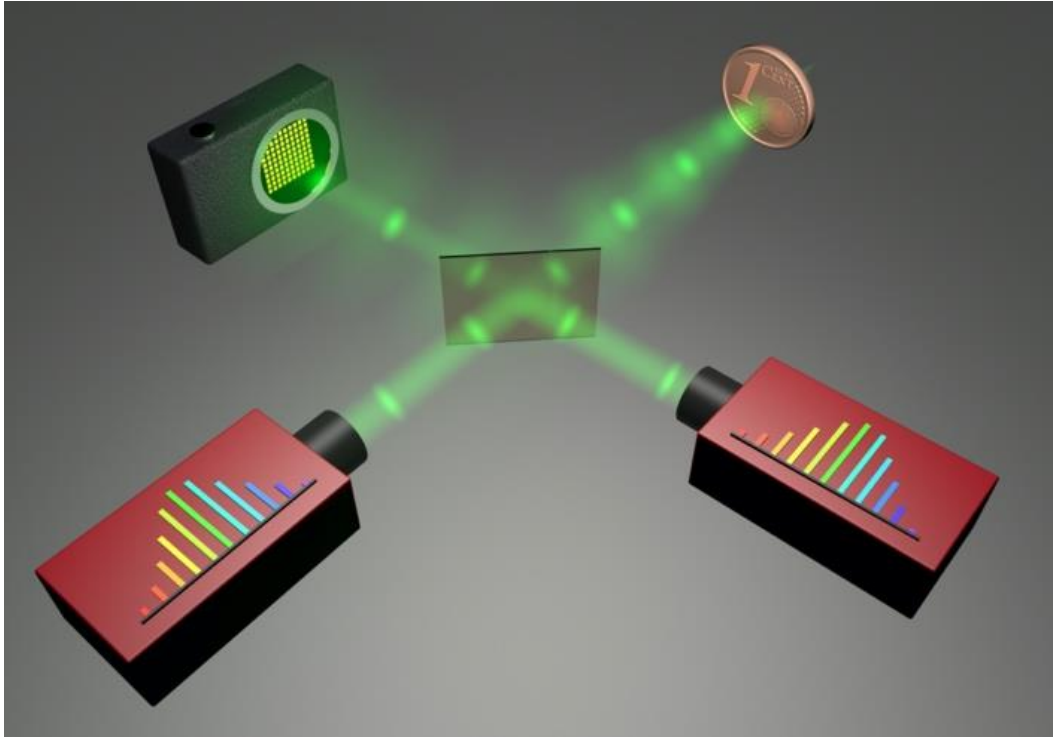

**Fig. 1. Dual-comb digital holography.** The regular train of pulses of a frequency-comb generator illuminates an object (here a coin in reflection). The wave scattered by the object spatially and temporally interferes with that of a reference comb at a lens-less detector matrix. Analysis of the temporal and spatial interference restores holograms and their reconstructed images (Fig.2).

As $f_{rep}$ is typically on the order of hundreds of MHz, the technique appears suited to large-scale objects. At the end of the recording, as many time-domain interferograms as there are pixels are obtained (Fig.2a). For each pixel, all spectral elements are acquired simultaneously by recording the interferogram. Each interferogram is Fourier transformed to reveal a complex spectrum of amplitude and phase (Fig.2b). Even with an in-line configuration, the holographic signal is spectrally separated from those of the non-interferometric zeroth-order signal and of the twin holograms (complex-conjugate duplicates of the object field), because it is mapped in a different frequency range. This straightforward way to eliminate the blurring zero-order light and the twin image can be seen as an analogous of phase-shifting holography [21]. Because the individual comb lines are resolved, as many holograms as there are comb lines are obtained. The amplitude and phase hologram at the well-defined optical frequency $nf_{rep}+f_{ceo}$ (where n is an integer and $f_{ceo}$ the carrier-envelope offset frequency) is mapped across the camera pixels at the radio-frequency $n\delta f_{rep}+\delta f_{ceo}$ (where $\delta f_{ceo}$ is the difference in carrier-envelope offset frequencies of the two combs). The set of complex holograms at all comb frequencies provides a hologram-hypercube (Fig.2c). The reconstruction is then similar to that used in other techniques of digital holography. For a Fresnel hologram, the hologram is multiplied by a chosen reference wave and an inverse Fresnel transform computes the back propagation at any depth of interest. Fig. 2d shows one amplitude image of a resolution target at an out-of-focus distance and the amplitude and phase maps at focus. Furthermore, the phase maps at different frequencies can be processed using well-established techniques of multi-wavelength holography. With dual-comb interferometry [13] though, a significant difference lies in the one-hundred-thousand frequencies, that can potentially be simultaneously measured. Hierarchical phase unwrapping [22] could for instance be implemented on an unprecedented scale to extend the ambiguity range of the quantitative phase derivations, from the wavelength of one comb line to that corresponding to the comb repetition frequency, while preserving low noise. Moreover, another asset of the highly frequency-multiplexed character of dual-comb holography is the possibility of multimodal diagnostics. For instance, if the object absorbs or if an absorber is in the beam path, the amplitude of the reconstructed images additionally reveals its spectral absorption map, enabling identification and quantification. Unlike dual-comb hyperspectral imaging [17-20], dual-comb digital holography provides 4D information (3D spatial and spectral dimensions).

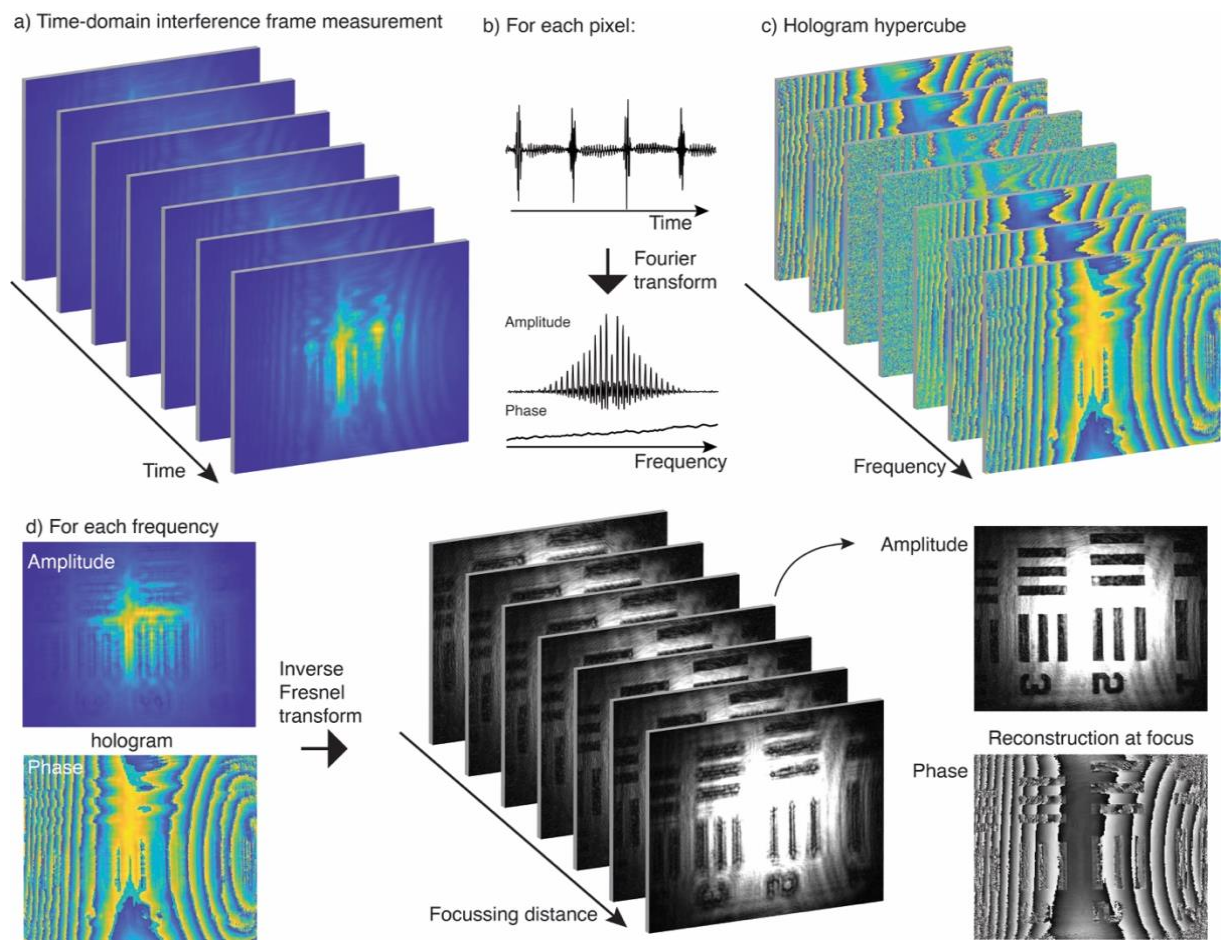

**Fig. 2. Reconstruction of dual-comb holographic amplitude and phase maps: a pictorial report using experimental data.** **(a)** The frames of the interference between the object and reference waves are sampled as a function of time on a detector matrix. The object is a resolution test target observed in transmission. **(b)** As many time-domain interferograms as there are pixels are measured. Each is Fourier transform to reveal frequency-domain amplitude and phase spectra with resolved comb lines. **(c)** The hologram hypercube includes as many complex holograms as there are comb lines: at a given frequency, the hologram includes the amplitude and phase of the spectrum across all detector pixels. **(d)** At a given frequency, an inverse Fresnel transform of the holograms generates amplitude and phase images over the entire range of focal distances and selected focused images are shown in the paper. The multiple phase maps will enable to unwrap the phase and to render the three-dimensional structure of the object (Fig.5).

Beside hyperspectral imaging, our technique might also be reminiscent -to readers familiar with frequency combs- of the single-photodetector dual-comb ranging technique which combines time-of-flight (LIDAR) and interferometric distance measurements [23]. Holography and laser ranging are however significantly different in their principle and in their applications. These differences apply equally to dual-comb implementations. In particular, LIDAR relies on the assumption that there is an unambiguous pointwise correspondence between the object and the detector pixel. LIDAR approaches that exploit a detector matrix (or multiplexing techniques) parallelize this hypothesis. Quite the opposite, in holography, the complex object wave is scattered and spatially interferes with the reference wave. The reconstruction of the hologram analyzes this two-dimensional interference pattern to restore the complex object field, with its depth and parallax. Uniquely, digital holography is a powerful lens-less and scan-free three-dimensional imaging method through linear and nonlinear [2, 3] media at the micro- and nano-scales, possibly at high temporal resolution [4]. Its applications [5, 11, 12] include optical contouring and deformation measurements, wavefront sensing, or 3D profiling over relatively short distances (compared to the hundreds of kilometers demonstrated by LIDAR techniques), microscopy and nanoscopy for life science [6, 7], particle imaging velocimetry, tomography, laser speckle contrast imaging, as well as the generation of complex 3D wavefronts through computer-generated-holograms in fields such as optogenetics [8], data storage [9], or near-eye display for virtual and augmented reality [10].

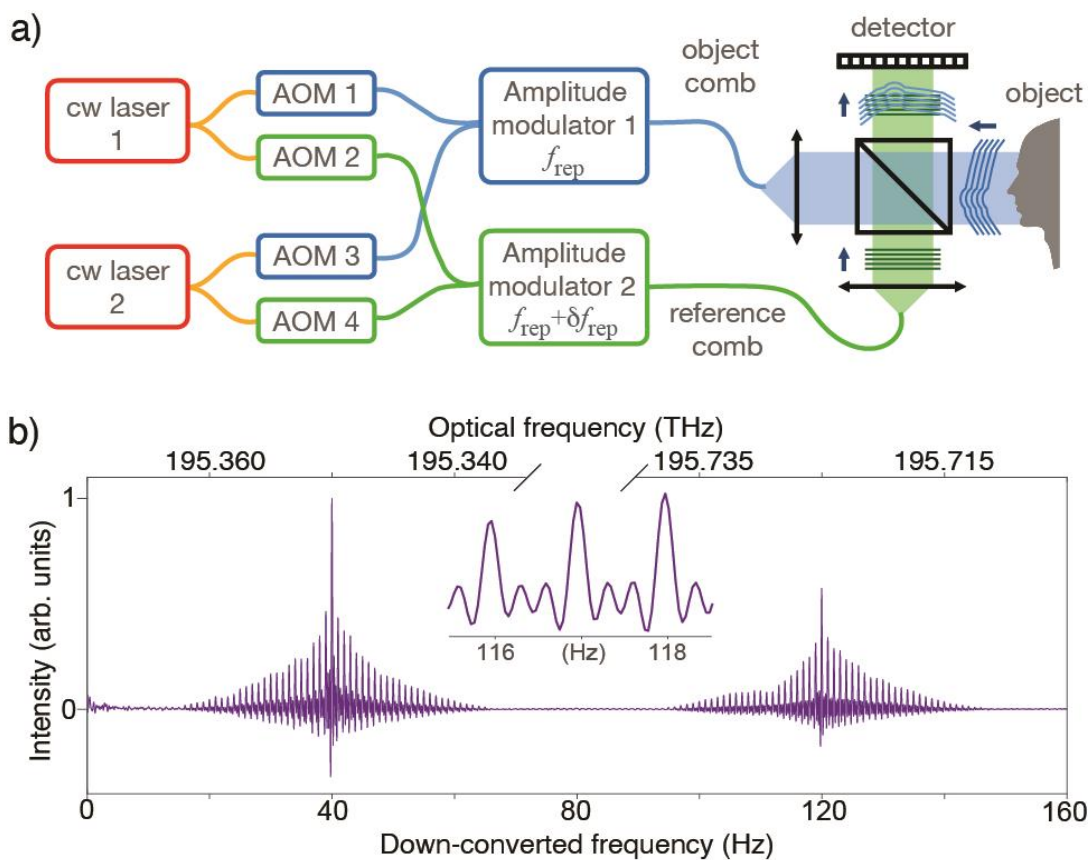

**Fig. 3. Dual-comb interferometry for inline Fresnel holography. (a)** Set-up for implementation with electro-optic modulators. Two continuous-wave lasers feed two amplitude modulators driven at slightly different repetition frequencies. Per modulator, two combs, of the same line spacing but with independently adjustable center frequencies, are generated. The object comb wave is scattered by the object and interferes at a detector matrix with the reference comb wave. **(b)** Dual-comb amplitude spectrum of pixel (160,128) of about 100 lines, with inset showing three individual comb lines. The amplitude and phase of the complex spectrum at one comb line position across all pixels generates the complex hologram at a given frequency.

We perform an experimental proof-of-principle of in-line Fresnel dual-comb holography in the near-infrared telecommunication spectral region (Fig. 3a, Methods). Our slow InGaAs detector matrix, of 320-Hz frame rate, dictates a small number of comb lines and long measurement times and orients the choice of a dual-comb interferometer based on electro-optic frequency comb generators. We generate two frequency combs, of slightly different repetition frequencies $f_{rep}$ (object comb) and $f_{rep}+\delta f_{rep}$ (reference comb), by dividing the output of a continuous-wave laser with a beam splitter, and sending each beam through an electro-optic amplitude modulator producing pulses of about 50 ps. We actually achieve broader spectral coverage by starting with two independent continuous-wave lasers of different frequencies [24] (here around 195 THz). Their outputs are superimposed so that each modulator produces two spectrally-separated frequency combs. These two combs do not need to be mutually coherent and no care is taken to achieve this. Each of the four modulator input beams is fine-tuned with an acousto-optic frequency shifter, enabling to map the two dual-comb spectra within the camera sampling bandwidth, at 40 Hz and 120 Hz respectively. The scattered waves are combined on a beam-splitter with the beam of the reference-comb generator. The object and reference waves beat on the lens-less InGaAs detector matrix of 320x256 pixels.

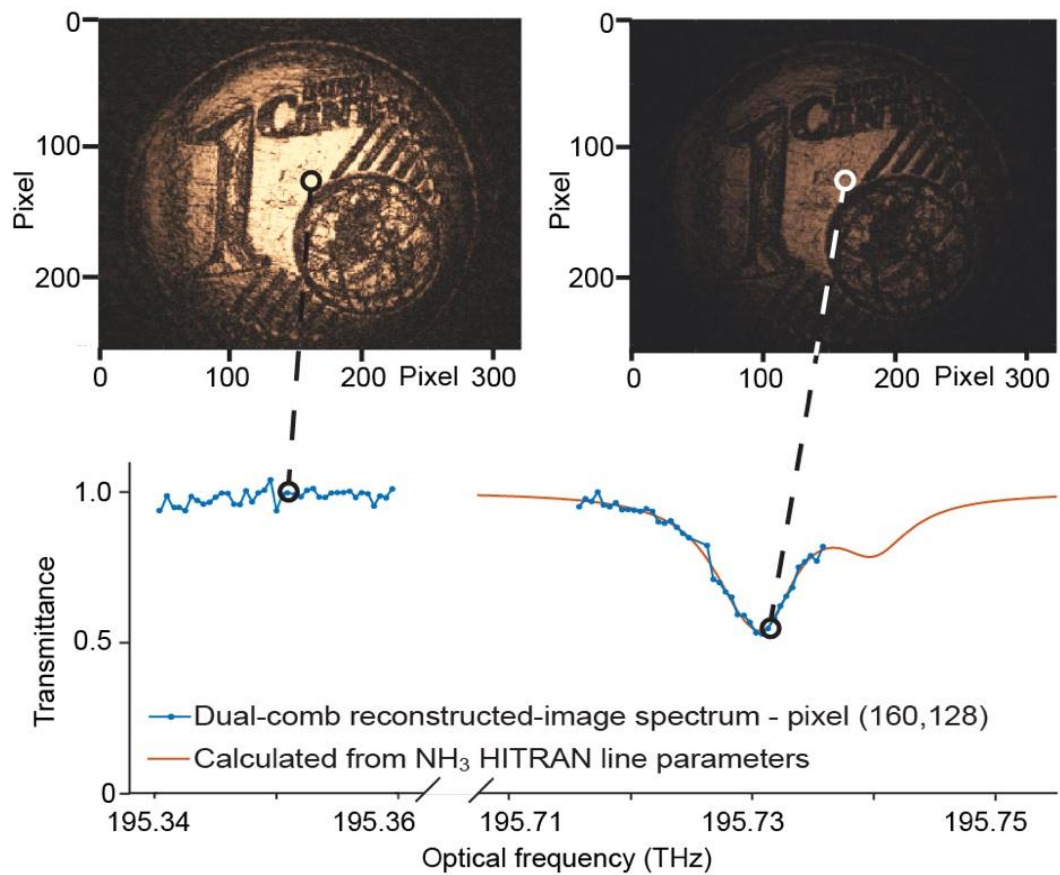

**Fig. 4. Illustration of the reconstruction of an experimental dual-comb hologram hypercube in presence of an absorber (gaseous ammonia) on the object beam path.** The object is a coin. (a) Two reconstructed amplitude images of the coin at focus (697 mm) for the frequencies of 195.351 THz and 195.731 THz, respectively. Each of the 81920 pixels of the reconstructed amplitude-image cube may be plotted as a high-resolution spectrum, such as that plotted here for pixel (160;128). The spectrum reveals the blended transition $PP(5, 3)a$ of the $\nu_1+\nu_3$ band in $^{14}NH_3$ [25], measured with sufficiently high signal-to-noise ratio to enable precise concentration measurements.

In a first validation experiment, the object is a resolution test target in transmission (Methods). The interferogram-hypercube is computed to a hologram-hypercube and then to reconstructed amplitude and phase maps, following the general procedure explained in Fig 2.

A second experiment illustrates the capabilities of our technique for multimodal diagnostics and for 3D reconstruction. The repetition frequency of the object comb and the difference in repetition frequencies are $f_{rep}$ =500 MHz and $\delta f_{rep}$ =1 Hz, respectively. The two continuous-wave lasers emit at the optical frequencies of 195.353 THz and 195.725 THz, respectively. The object beam is reflected and scattered by a coin (1 Euro-cent) and passes through 40 cm of ammonia vapor in air along each way. An interferogram-hypercube, consisting of 81920 pixel-interferograms, is measured over a time of 91 seconds. The Fourier transform of the interferograms leads to amplitude (Fig.3b) and phase spectra with 100 resolved comb lines for each pixel. Each dual-comb hologram-spectrum shows two down-converted combs, centered at 40 Hz (for the combs centered at 195.353 THz) and 120 Hz (for the combs centered at 195.725 THz) respectively. Dual-comb holography offers high-quality complex images at any focusing distance: an inverse Fresnel transform, assuming spherical waves, computes a set of amplitude and phase images (Supplementary Fig.S1a) for 100 different optical frequencies, at any desired depth. At the focus for the coin -a focusing distance of 69.7 cm-, two amplitude images, one outside absorption and one within an $NH_3$ absorption line, show excellent agreement (Fig. 4), but as expected, the latter is significantly attenuated. For each of the 81,920 pixels, the amplitude variations across the comb-line frequencies provide the absorption along the beam path. The $NH_3$ absorption spectrum of pixel (160;128) shows good agreement with a spectrum adjusted from HITRAN line parameters (Fig.4), enabling concentration measurements. As usual in digital holography, the images can be reconstructed even in presence of faulty detector pixels (Supplementary Fig.S1). The benefits of the large frequency-multiplexing are not limited to the possibility of multimodal diagnostics. The ambiguity range of the holographic measurements is also greatly enhanced. Using multi-frequency phase unwrapping, limited to the phase images of the comb lines which do not experience absorption by the sample, a quantitative three-dimensional phase map is reconstructed (Fig.5).

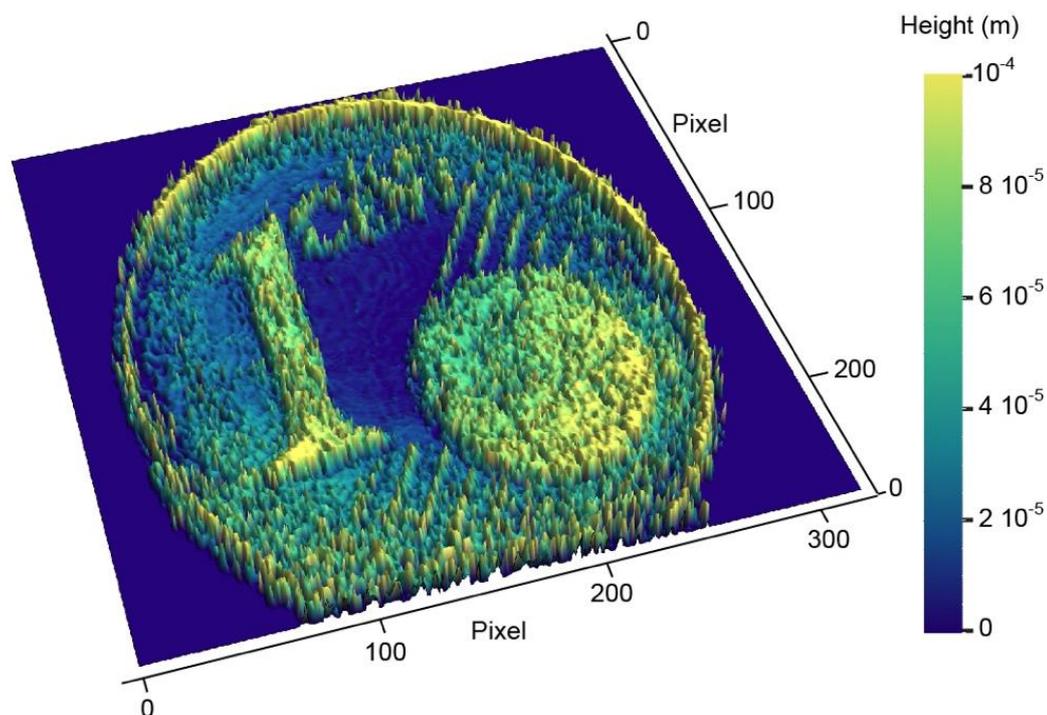

**Fig. 5. Reconstructed three-dimensional image of the phase map**, using about 60 comb lines at different optical frequencies and multi-frequency phase unwrapping.

Future work will explore the precision and accuracy of our technique. Faster cameras with a higher number of pixels and a larger sensor are available in the visible spectral range. They will dramatically increase the number of comb lines, the measurement speed and the spatial resolution, enabling powerful implementations with self-referenced fiber-based frequency comb synthesizers. Dual-comb interferometers already show an intriguing potential in spectroscopy [13], in spectro-imaging [17, 26] and in ranging [23]. Their unique combination of broad-spectral-bandwidth, long temporal coherence and multi-heterodyne read-out offers holography a distinctive host of powerful features – frequency multiplexing, accuracy, precision, speed, large ambiguity range –, likely to conquer new frontiers in scan-free wavefront reconstruction and three-dimensional metrology.

## Methods

**Dual-comb interferometer for proof-of-principle demonstration of inline Fresnel holography.**

The reported demonstration of dual-comb hyperspectral digital holography makes expedient use of equipment already available in our laboratory. The experimental choices are pragmatically based on the limitations of these tools. At the end of the Methods section, we sketch a short outlook of the opportunities with optimized design.

The main limitation for our experiment in the near-infrared telecommunications region is imposed by the slow 320-Hz frame rate of the detector matrix, a lens-less InGaAs thermo-electrically-cooled camera of 320x256 pixels. At such slow frame rates, it is difficult to accommodate a very large number of comb lines in dual-comb interferometry, and it is challenging to maintain the mutual coherence of two frequency combs over the necessary long observation times. We therefore resort to modulator-based combs with fewer than 100 comb lines, but excellent passive mutual coherence [27, 28]. The few previous experiments combining detector matrices and dual-comb interferometers have all made use of a double acousto-optic-frequency-shifter scheme [17-20], where the two combs are shifted by a slightly different radio-frequency, enabling to center the down-converted comb at frequencies on the order of tens of Hz. To increase the effective spectral span, we actually employ two sub-combs in each frequency comb, separated in frequency by several hundred GHz, as has been done before in spectrally-tailored dual-comb spectroscopy with thin-film lithium-niobate-on-insulator electro-optic micro-rings [24]. By acousto-optically fine-tuning the carrier frequencies of each of the four combs, we can freely shift the dual-comb interference frequencies so that the signals from both down-converted sub-combs fall within the detector-matrix bandwidth and aliasing is avoided.

As shown in Fig. 3a, we start with two fiber-coupled broadly tunable free-running continuous-wave lasers with narrow linewidths, emitting around 195 THz. Each continuous-wave laser beam is split into two beams, each of which is sent into an acousto-optic modulator (AOM) which acts as a frequency shifter. Modulator AOMi (with i=1,2,3,4) is driven at the radio-frequency $\delta f_i$, with $\delta f_1$=25.0 MHz, $\delta f_2$= 25.0 MHz+40 Hz, $\delta f_3$= 40.0 MHz, $\delta f_4$= 40.0 MHz+120 Hz. The beams at frequencies $f_1+\delta f_1$ and $f_2+\delta f_3$ are combined and are sent into an electro-optic amplitude modulator (Amplitude Modulator 1). Driven by a synthesizer and a pulse generator, Amplitude Modulator 1 generates a train of 50-ps pulses at the radio-frequency $f_{rep}$=1000 MHz (Fig.2) or 500 MHz (Fig.3-5). Similarly, the beams at frequencies $f_1+\delta f_2$ and $f_2+\delta f_4$ are sent into Amplitude Modulator 2. Amplitude Modulator 2 also generates a train of 50-ps pulses at the slightly different radio-frequency $f_{rep}+\delta f_{rep}$ =1000 MHz + 2 Hz (Fig.2) or 500 MHz+1Hz (Fig.3-5). The fiber-coupled amplitude modulators have a frequency bandwidth of 20 GHz and an extinction ratio of 40 dB at 1550 nm. At the output of Amplitude Modulator 1, two spectrally separated sub-combs are generated, centered at the optical frequencies $f_1+\delta f_1$ and $f_2+\delta f_3$, respectively. These two sub-combs, which form the object comb, have a line spacing equal to $f_{rep}$ and count about 50 lines each when $f_{rep}$ =500 MHz (about 25 lines when $f_{rep}$ =1000 MHz). As the two sub-combs are generated from two independent free-running continuous-wave lasers, they are expected to show poor mutual coherence, which does not matter in our experiment. The output of Amplitude Modulator 2 is similar, with its corresponding parameters, and generates the reference comb with two sub-combs.

At the output of the amplitude modulators, the beams of the object and reference combs are collimated in free space to a diameter which is slightly larger than that of the detector matrix. The beam of the object comb is transmitted through or reflected and scattered onto an object. The object and reference waves are then combined on a beam-splitter cube and beat on a detector matrix sensor. The pixel size of the sensor is 30x30 µm$^2$ and its dynamic range is 14 bits. A frame-grabber card captures the sensor frames and streams them to the hard-drive of a computer.

Two down-converted sub-combs of a line spacing $\delta f_{rep}$ are generated in the detector signal: one sub-comb centered at $\delta f_2-\delta f_1$ = 40 Hz, the other sub-comb centered at $\delta f_4-\delta f_3$ = 120 Hz. For each matrix pixel, the time-domain interferogram is Fourier transformed and further processing is as described in Fig.2. The spectra confirm that the optical sub-combs, generated from the same continuous-wave laser and different amplitude modulators, show excellent mutual coherence, here longer than 10 seconds, owing to the shared optical carrier.

**Recording conditions.**
For the data displayed in Fig. 2, the two continuous-wave lasers emit at the respective frequencies of $f_1$= 194.135 THz and $f_2$=195.405 THz. The object-comb line spacing is $f_{rep}$=1000 MHz and the difference in repetition frequencies is $\delta f_{rep}$ = 2 Hz. Each comb includes about 25 lines (12-13 lines per sub-comb). The object is a resolution test target (1951 USAF) which works in transmission. An interferometric hypercube, consisting of 81920 interferograms, is measured within a total measurement time of 3.5 seconds.

In Fig. 4 and Fig. 5, the two continuous-wave lasers emit at the respective frequencies of $f_1$=195.353 THz and $f_2$=195.725 THz. The object is a 1 €-cent coin, which scatters the reflected light. Before and after the reflection onto the coin, the object waves go through a box, with holes to let the light in and out on each side and ammonia water at the bottom, and through a telescope made of two spherical mirrors that magnifies the incoming object beam about two-fold before it reaches the coin (and demagnifies the exiting object waves of the same factor). An interferogram-hypercube, consisting of 81920 interferograms, is measured within 7 seconds. Thirteen hypercubes are averaged, leading to the total measurement time of 91 seconds.

**Outlook.**
Our technique can be dramatically improved if it is moved to the range of visible and near-infrared wavelengths down to 1 µm. Much faster cameras, based on silicon technology, are commercially available, with rates higher than 3 10$^5$ frames per second at 600x400-pixel size. Together with comb generators based on mode-locked fiber lasers, they will make it possible to use much broader spectral spans and to acquire dual-comb hypercubes much faster than in the reported first proof-of-principle experiments. An interferometer of high mutual-coherence based on two frequency-comb synthesizers fully referenced to a radio-frequency clock, as conveniently available from fiber-laser systems [13], will be key to exploring the frontiers in precision and accuracy of our technique and will enable dual-comb holography to evolve into a tool of dimensional metrology.

**Acknowledgments.** We thank Karl Linner for technical support. Funding from the Max-Planck Society, the Carl-Friedrich von Siemens Foundation, the Max-Planck Fraunhofer cooperation program, the China Scholarship Council is gratefully acknowledged.

## Supplementary Information: Supplementary Figure S1

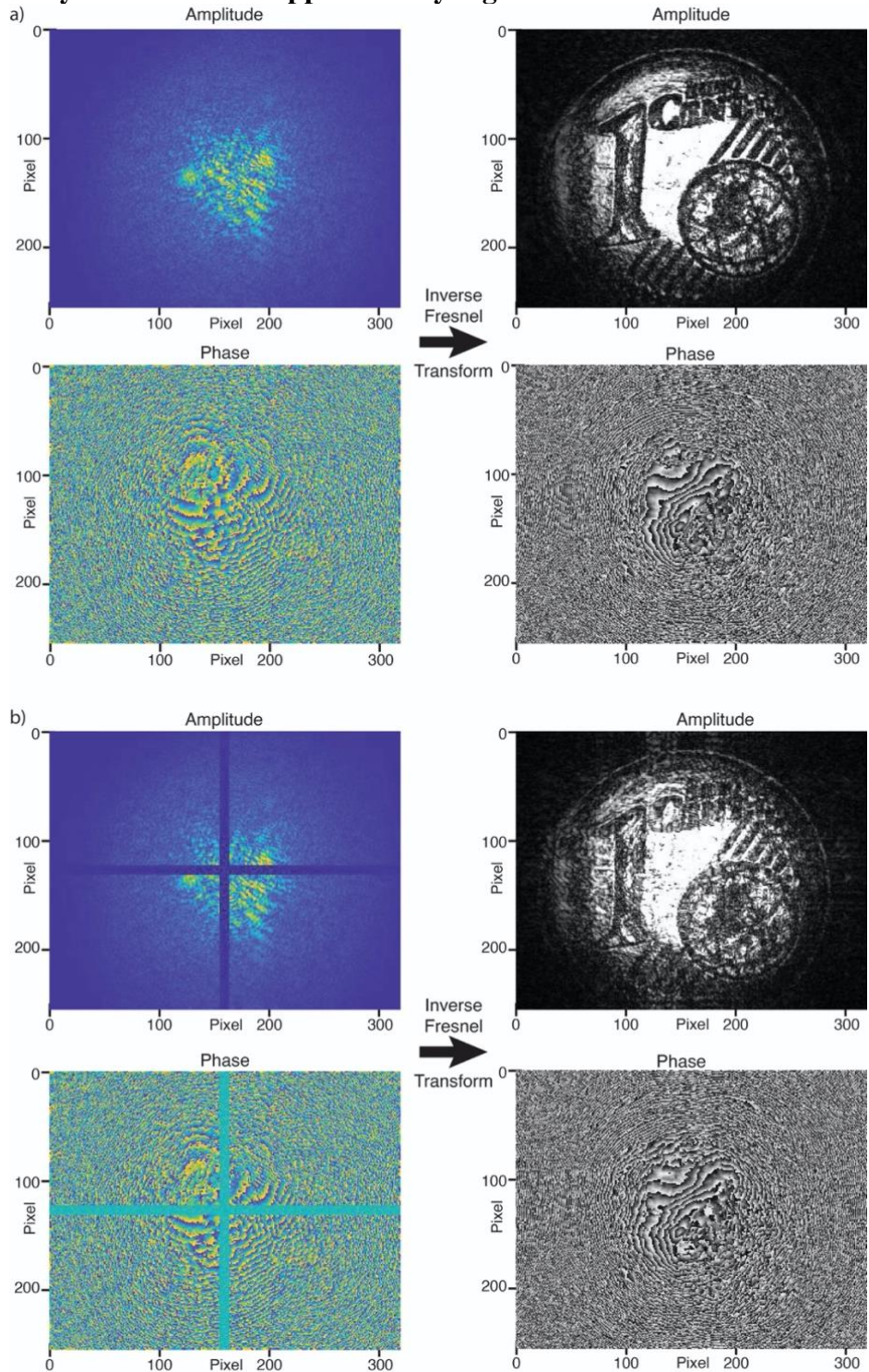

**Supplementary Figure S1. Robustness of dual-comb holography against missing pixels.**
**(a)** The complex (amplitude and phase) hologram at 195.351 THz is inversed Fourier transformed. At a focusing distance of 0.697 m, the reconstructed amplitude and phase images of a 1 €-cent coin are at focus.
**(b)** The same complex hologram as in (a) is truncated: The samples for 10 entire horizontal pixel lines and 10 entire vertical pixel lines, in the middle of the detector matrix, are arbitrarily set to zero to provide a drastic simulation of the effect of faulty pixels. Nonetheless, the reconstructed images (also at 0.697 m) still provide a recognizable representation of the coin. This robustness is a well-known characteristic of the digital holography technique.